# The Isgur-Wise Function

R.D. Kenway

Department of Physics & Astronomy, The University of Edinburgh,
The King's Buildings, Edinburgh EH9 3JZ, Scotland

After a brief introduction to the Heavy-Quark Effective Theory (HQET), I review the extraction of the Isgur-Wise function from lattice QCD calculations of the matrix elements for semi-leptonic decays of heavy-light pseudoscalar mesons both into pseudoscalar and into vector mesons. This work is beginning to test the heavy-quark spin-flavour symmetries around the charm mass and to indicate the size of $O(1/m_c)$ corrections. An alternative approach to put the HQET on the lattice offers the prospect of computing the Isgur-Wise function directly.

## 1. INTRODUCTION

### 1.1. HQET

The motivation behind Isgur-Wise function calculations is to be able to extract CKM matrix elements from experimental measurements of heavy-quark decays in a model-independent way. Of particular interest are the semi-leptonic decays $\bar{B} \to D\ell\bar{\nu}$ and $\bar{B} \to D^*\ell\bar{\nu}$, from which it is possible to extract $V_{cb}$.

I shall consider only heavy-light mesons in which the heavy-quark mass, $m_Q$, is much greater than $\Lambda_{\rm QCD}$. Then the four-velocity of the heavy quark is essentially the same as the four-velocity of the meson and, in its rest frame, the heavy quark appears to the light degrees of freedom as a static colour source. The spin interaction between the heavy quark and the light degrees of freedom is suppressed by the heavy-quark mass. So, if a weak current changes the heavy quark into a heavy quark of a different flavour (mass), or flips its spin, without changing its velocity, then the light degrees of freedom see no change. This is the reason for the spin-flavour symmetries of the HQET. If the weak current changes the velocity of the heavy quark, then the light degrees of freedom have to adjust themselves, and this results in a form-factor suppression. The overlap between the state of the light degrees of freedom in the presence of a colour source moving with velocity $v$, and that in the presence of a colour source moving with velocity $v'$, is the Isgur-Wise function [1]. This is a fundamental quantity in QCD, which can only be determined non-perturbatively.

Since the velocity of the heavy quark is conserved by soft QCD processes, it may be used to project out the upper spinor components of the heavy-quark field, $Q$,

$$\frac{1+\slashed{v}}{2} Q = e^{-im_Q v \cdot x} H, \qquad \slashed{v} H = H. \tag{1}$$

The HQET is obtained by integrating out the lower components [2]. There is a different field $H$ for each velocity [3]. A semi-leptonic decay involves a transition between a heavy quark with velocity $v$ and another with velocity $v'$ so, at leading order in $1/m_Q$, the Lagrangian is

$$\mathcal{L}_{\rm HQET} = \bar{H} iv \cdot DH + \bar{H}' iv' \cdot DH'. \tag{2}$$

### 1.2. Semi-leptonic form factors

The matrix elements of the vector and axial vector currents between pseudoscalar ($P$) and vector ($P^*$) meson states may be expressed in terms of six form factors, $h_i = h_i(\omega)$:

$$\langle P', v'|V_\mu|P,v\rangle = (v+v')_\mu h_+ + (v-v')_\mu h_- \tag{3}$$

$$\langle P^*, v', \varepsilon|V_\mu|P,v\rangle = i\epsilon_{\mu\nu\rho\sigma}\varepsilon^{*\nu}v'^\rho v^\sigma h_V \tag{4}$$

$$\langle P^*, v', \varepsilon|A_\mu|P,v\rangle = (\omega+1)\varepsilon^*_\mu h_{A_1}$$
$$-\varepsilon^* \cdot v[v_\mu h_{A_2} + v'_\mu h_{A_3}] \tag{5}$$

where $v$ and $v'$ are the velocities of the incoming and outgoing mesons, $\omega = v \cdot v'$, and I have used a mass-independent normalisation for the states. In the HQET, the spin-flavour symmetry imposes constraints on the form factors. Eq. (1) gives

$$(v - v')_\mu \bar{H}' \gamma^\mu H = 0 \quad \Rightarrow \quad h_-(\omega) = 0. \quad (6)$$

The time component of the vector current generates the flavour symmetry:

$$\langle P', v | \int d^3 x \, \bar{H}' \gamma^0 H | P, v \rangle = \langle P', v | P', v \rangle$$
$$\Rightarrow h_+(1) = 1. \quad (7)$$

Flipping the spin of the heavy quark changes a pseudoscalar meson into a vector meson. The full consequences of the heavy-quark symmetry are [4]

$$\langle M', v' | \bar{H}' \Gamma H | M, v \rangle = -\xi(\omega) \text{Tr}[\bar{\mathcal{M}}' \Gamma \mathcal{M}] \quad (8)$$

$$\mathcal{M} = \frac{1 + \slashed{v}}{2} \begin{cases} -\gamma_5 & \text{for } |P, v\rangle \\ \slashed{\varepsilon} & \text{for } |P^*, v, \varepsilon\rangle \end{cases} \quad (9)$$

so that, in the limit $m_Q \to \infty$,

$$h_+ = h_V = h_{A_1} = h_{A_3} = \xi(\omega), \quad \xi(1) = 1, \quad (10)$$

$$h_- = h_{A_2} = 0. \quad (11)$$

$\xi(\omega)$ is the Isgur-Wise function. It is normalised to one at zero velocity transfer (zero recoil).

### 1.3. Renormalised Isgur-Wise function

The above results are only exact in the limit of exact spin-flavour symmetry. We want the corresponding results in QCD, when the heavy-quark mass is large but finite. QCD and the HQET contain the same long-distance physics, but differ at short distances, where QCD resolves the dynamics of the heavy quark through the high momentum that can flow in internal loops. These short-distance corrections are included by expressing a current in QCD as an expansion in terms of local operators in the HQET:

$$J^{\text{QCD}} \simeq \sum_i C_i(\mu) J_i(\mu)^{\text{HQET}} + O(1/m_Q). \quad (12)$$

Although in QCD the vector and axial-vector currents are not renormalised, the corresponding HQET currents are not symmetry currents and so depend on the renormalisation scale $\mu$. In Eq. (12) this dependence is cancelled by that of the Wilson coefficients, $C_i$.

The Wilson coefficients have been calculated to next-to-leading order by Neubert [5]. This calculation is complicated by the fact that there are two heavy-quark scales: $m_b$ and $m_c$. Neubert's review [6] provides a complete discussion. He shows that the Wilson coefficients factorise:

$$C_i(\omega, \mu) = \hat{C}_i(m_b, m_c, \omega) K(\omega, \mu) \quad (13)$$

where $K$ contains all the dependence on $\mu$, is independent of the spin structure, and

$$K(1, \mu) = 1. \quad (14)$$

Therefore, up to corrections of $O(1/m_Q)$, the matrix element of a heavy-quark current in QCD is given in terms of HQET matrix elements by

$$\langle M', v' | J | M, v \rangle = \sum_i C_i \langle M', v' | \bar{H}' \Gamma_i H | M, v \rangle$$
$$= -\xi_{\text{ren}} \sum_i \hat{C}_i \text{Tr}[\bar{\mathcal{M}}' \Gamma_i \mathcal{M}], \quad (15)$$

where

$$\xi_{\text{ren}}(\omega) = \xi(\omega, \mu) K(\omega, \mu) \quad (16)$$

is the renormalisation-group invariant Isgur-Wise function, which is also normalised to one at zero recoil, because of Eq. (14).

The short-distance corrections break the heavy-quark symmetry so that, from Eq. (15), the naive relations between the form factors, Eq. (10) and (11), are replaced by

$$h_+ = \left[ \hat{C}_1 + \frac{\omega + 1}{2} (\hat{C}_2 + \hat{C}_3) \right] \xi_{\text{ren}} \quad (17)$$

$$h_- = \frac{\omega + 1}{2} (\hat{C}_2 - \hat{C}_3) \, \xi_{\text{ren}} \quad (18)$$

$$h_V = \hat{C}_1 \, \xi_{\text{ren}} \quad (19)$$

$$h_{A_1} = \hat{C}_1^5 \, \xi_{\text{ren}} \quad (20)$$

$$h_{A_2} = \hat{C}_2^5 \, \xi_{\text{ren}} \quad (21)$$

$$h_{A_3} = (\hat{C}_1^5 + \hat{C}_3^5) \, \xi_{\text{ren}}. \quad (22)$$

Neubert [5,6] gives expressions for $\hat{C}_i$ and $\hat{C}_i^5$ as functions of $\omega$, $m_b$ and $m_c$. He finds that, for the physical $b$ and $c$ masses, the next-to-leading-order corrections can be as large as 15%. This result is important, because we extract estimates of $\xi_{\text{ren}}$ from measurements of the form factors in QCD using Eq. (17)-(22), i.e.,

$$\xi_{\rm ren} \sim \frac{h}{\hat{C}} \equiv h_{\rm rad\ corr}. \qquad (23)$$

It is useful to parametrise the Isgur-Wise function by means of its slope at zero recoil, $\xi'(1) = -\rho^2$. Most models agree in the kinematically accessible region, $1 < \omega < 1.6$, and so I will define $\rho^2$ using the Bauer-Stech-Wirbel (BSW) model:

$$\xi_\rho(\omega) = \frac{2}{\omega+1} \exp\left[-(2\rho^2-1)\frac{\omega-1}{\omega+1}\right]. \qquad (24)$$

QCD sum rules give bounds on $\rho^2$ [6]:

Bjorken: $\quad \rho^2 > 0.25 \qquad (25)$

Voloshin: $\quad \rho^2 < 0.25 + \dfrac{m_M - m_Q}{E_{\min}} \sim 1. \qquad (26)$

### 1.4. Power corrections

At order $1/m_Q$ there are new universal functions (see [6]). Lattice calculations are not yet precise enough to disentangle these. The most useful statement is Luke's Theorem [7]: $O(1/m_Q)$ corrections to meson-decay matrix elements vanish at zero recoil. This only protects

$$h_+(1) = \sum_i \hat{C}_i(1) + O(1/m_Q^2) \qquad (27)$$

$$h_{A_1}(1) = \hat{C}_1^5(1) + O(1/m_Q^2) \qquad (28)$$

because the other form factors are multiplied by terms which vanish anyway at $\omega = 1$. Using QCD sum rules for $b \to c$, Neubert [6] estimates the power corrections to be around 3% for $h_+$ and $h_{A_1}$, throughout the kinematic range. In contrast, the correction to the relation between $h_V$ and $\xi$ is around 30%. A challenge for lattice calculations, with important phenomenological implications, is the determination of ratios such as $h_V/h_{A_1}$ and $h_+/h_{A_1}$.

## 2. RESULTS FROM LATTICE QCD

Bernard et al. [8,9] were the first to report results for the Isgur-Wise function extracted from form factors computed in quenched lattice QCD. UKQCD has now also reported results [10]. Both groups have presented further analyses at this conference [11–15]. The parameter values used are given in Table 1. Both groups use heavy-quark masses around charm and light-quark masses around strange. UKQCD uses the $O(a)$-improved Sheikholeslami-Wohlert (SW) fermion action [16], whereas Bernard et al. use the Wilson fermion action.

Table 1
QCD simulations using $16^3$ and $24^3$ lattices

| $\beta$ | $a^{-1}$(GeV) | $m_P$(GeV) | $m_{\bar{q}q}$(GeV) |
|---|---|---|---|
| Bernard et al: Wilson action, | | | $\sim$ 20 configs |
| 6.0 | 2.0 | 1.6–2.3 | 0.6–1.0 |
| 6.3 | 3.2 | 1.6–2.7 | 0.6–1.0 |
| UKQCD: SW action, | | 36 and 60 configs | |
| 6.0 | 2.0 | 1.6–2.1 | 0.5–0.8 |
| 6.2 | 2.7 | 1.5–2.4 | 0.5–0.8 |

### 2.1. Pseudoscalar form factors: $P \to P'\ell\bar{\nu}$

For elastic scattering ($m_Q = m_{Q'}$), in QCD, because of current conservation,

$$h_-^{\rm elastic}(\omega) = 0, \quad h_+^{\rm elastic}(1) = 1 \qquad (29)$$

to all orders in $1/m_Q$. It follows that, close to zero recoil, the $O(1/m_Q)$ differences between $h_+$ and $\xi$ should be small, and that $h_{+{\rm rad\ corr}}^{\rm elastic}$ should provide a good determination of $\xi_{\rm ren}$. Assuming that heavy-quark symmetry applies to $b \to c$ transitions, this opens the possibility of a model-independent determination [6] of $V_{cb}$ from experimental measurement of

$$\frac{d\Gamma(\bar{B} \to D^*\ell\bar{\nu})}{d\omega} = \text{known factor} \times \xi^2(\omega)|V_{cb}|^2. (30)$$

The matrix elements of the lattice vector current between pseudoscalar states of various momenta are obtained in the usual way from 3-point correlation functions, using amplitudes and energies from fits to 2-point functions. To fix the normalisation of the lattice vector current ($Z_V V_\mu^{\rm latt} = V_\mu$), both groups divide by the forward matrix element of $V_0$ at zero momentum, which in the continuum is 2. This cancels $Z_V$ and any field normalisation, e.g., as is required for Wilson fermions at large quark mass [9]. This approach should tend to reduce the effect of lattice artefacts which render the normalisations uncertain. UKQCD data for the SW action show that $Z_V$ is insensitive to the light-quark mass, or to whether momentum 0 or $\pi/12$ is used, but it does depend on the heavy-quark mass. Thus, the procedure is unambiguous for elastic matrix elements, given that $h_+$ is normalised at zero recoil.

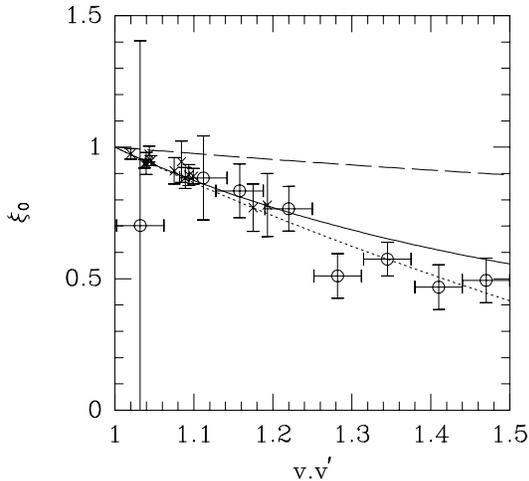

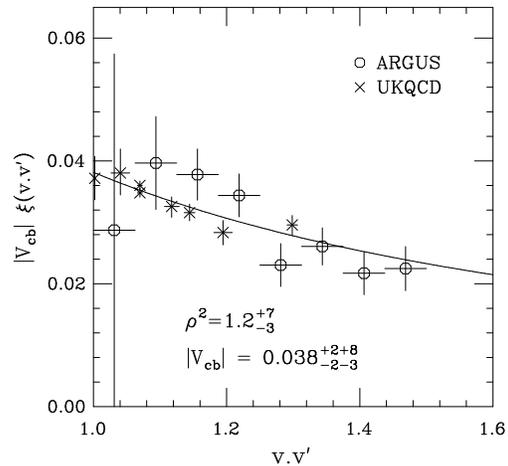

Figure 1. Isgur-Wise function obtained by Bernard et al. from $h_+$ (crosses) compared with ARGUS data (circles).

Figure 2. Isgur-Wise function obtained by UKQCD from $h_+$, compared with ARGUS data.

The estimate of the Isgur-Wise function obtained by Bernard et al. [9,11] is shown in Figure 1, along with experimental data from ARGUS [17]. The lattice results are from a mixture of $\beta$ and $m_Q$ values, at fixed light-quark mass around $m_s$. No radiative corrections are applied.

The corresponding results from UKQCD [10, 12] are shown in Figure 2. Here the data are at fixed $\beta$ (6.2), fixed heavy-quark mass (charm), have been chirally extrapolated, and have had radiative corrections applied as in Eq. (23). A fit of Eq. (24) to the lattice data gives

$$\rho^2 = \begin{cases} 1.4 \pm 2 \pm 4 & \text{Bernard et al.} \\ 1.2 ^{+\ 7}_{-\ 3} & \text{UKQCD} \end{cases} \quad (31)$$

where some estimate of systematic errors has been included by both groups. The corresponding BSW functions are shown as solid curves. Using these to extrapolate the experimental data to zero recoil gives

$$|V_{cb}| = \begin{cases} 0.044 \pm 7 \pm 5 & \text{Bernard et al.} \\ 0.038 ^{+\ 2}_{-\ 2} ^{+\ 8}_{-\ 3} & \text{UKQCD} \end{cases} \quad (32)$$

where the first error is experimental and the second is from the lattice (the $B$ lifetime is taken to be 1.49ps). Systematic errors apart, the lattice calculations are quite precise, in comparison with the experimental data, and the agreement between them is good.

I will now turn to the systematics. The form factor $h_+$ is obtained from the elastic matrix element by setting $h_- = 0$ (Eq. (29)). This is true in the continuum, but not necessarily so on the lattice. UKQCD has tested this [12] and finds that, at $\beta = 6.2$, the measured values of $h_-$ are consistent with zero to within $2\sigma$, provided that the maximum momentum in any external particle is less than or equal to $\sqrt{3}(\pi/12)$.

Also reported [12] are the results of a more refined analysis of the $\beta = 6.2$ UKQCD data, using a correlated fit to all non-zero matrix elements and more momentum combinations. This suggests that $\rho^2$ decreases slightly with light-quark mass to a value of $0.9^{+4}_{-3}$ in the chiral limit, although the data are consistent with a constant value around 1.2. Bernard et al. [9] also note this tendency, but their data are rather noisy.

At $\beta = 6.0$, UKQCD determines $h_+$ and $h_-$ at fixed $\omega > 1$ and at fixed $m_D$, as an expansion in the hopping parameter for the $b$ quark [13]. Although the slight dependence of $Z_V$ on $m_Q$ is ignored in this preliminary analysis, UKQCD finds that $h_{+\text{rad corr}}$ is approximately independent of $m_B$, so that power corrections are small, as expected, indicating that the heavy-quark flavour symmetry holds at around the charm mass (within large statistical errors), whereas, $h_-/h_+$ shows a stronger dependence on $m_B$ than the size of the short-distance and power corrections estimated by Neubert [6] would indicate. If systematic errors can be controlled, such measurements will be important predictions for the lattice.

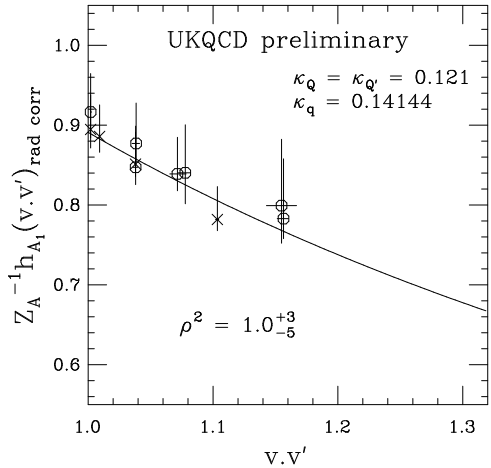

Figure 3. Isgur-Wise function, normalised to $Z_A^{-1}$, obtained from $h_{A_1}$ at fixed heavy-quark mass (circles), and from different heavy-quark masses at maximum recoil (crosses).

## 2.2. Vector and axial-vector form factors: $P \to P^* \ell \bar{\nu}$

So far, only UKQCD has reported results for the Isgur-Wise function extracted from matrix elements for the decay of a pseudoscalar into a vector meson [14,15]. This preliminary analysis is providing the first tests of the heavy-quark spin symmetry in semi-leptonic decays.

Although the lattice vector current may be normalised using $\langle P, v | V_\mu | P, v \rangle = 2v_\mu$, the normalisation of the axial-vector current, $Z_A A_\mu^{\text{latt}} = A_\mu$, must be determined separately. The UKQCD analysis assumes that $Z_A$ is an overall constant, and shows that this assumption is supported (within relatively large statistical errors) by estimating $Z_A^{-1} = h_{A_1}(1)_{\text{rad corr}}$ [14]. The further uncertainty in the normalisation of the quark fields is much reduced by using the SW action [18].

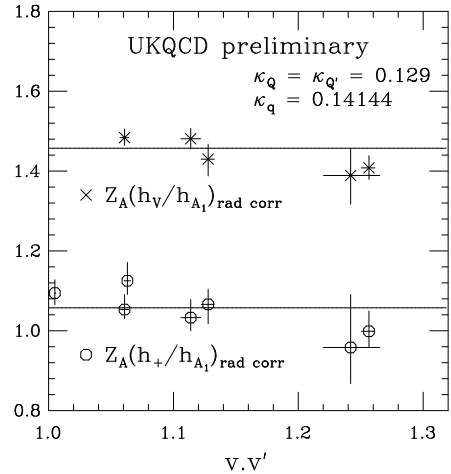

Figure 4. Ratios of estimates of the Isgur-Wise function.

Figure 3 [15] shows the Isgur-Wise function extracted from the axial-vector form factor. Data are included at fixed equal heavy-quark masses, $m_Q = m_{Q'}$, and various momenta, and at fixed momentum transfer ($q^2 = 0$) and various heavy-quark masses, $m_Q \neq m_{Q'}$, for which

$$\omega(q^2 = 0) = \frac{1}{2} \left( \frac{m_P}{m_{P^*}} + \frac{m_{P^*}}{m_P} \right). \qquad (33)$$

The agreement between the two sets of data provides further evidence for heavy-quark flavour symmetry at around the charm mass.

Evidence for spin symmetry at the charm mass is given by the fact that the ratio $h_+/h_{A_1}$ in Figure 4 is approximately constant and close to 1. Both form factors in this ratio are protected by Luke's Theorem. In contrast, the ratio of the vector to axial-vector form factors, also shown in Figure 4, is significantly different from 1, although roughly constant [15]. This supports the QCD sum rule result that $h_V$ suffers large $O(1/m_Q)$ corrections. These appear to be roughly proportional to $\xi$ and around 40% at the charm quark mass. The size of these corrections and their dependence on the heavy-quark masses [15] are roughly consistent with the QCD sum rule estimate of a 30% effect for $b \to c$ [6].

## 3. LATTICE HQET

The Isgur-Wise function may also be calculated by formulating the HQET, Eq. (2), directly on a lattice. Mandula and Ogilvie [19] solve a lattice version of the heavy-quark propagator equation,

$$iv \cdot D\tilde{S}(x,y;v) = \delta(x,y), \qquad (34)$$

numerically, by forward recursion, whereas Aglietti [20] expands $\tilde{S}$ as a power series in the velocity $v$, obtaining $\xi'(1)$ at $O(v^2)$.

At this conference, Mandula and Ogilvie presented results for $\xi(\omega)$, obtained by combining $\tilde{S}$ with Wilson light-quark propagators at $\beta = 5.7$ [21]. They encounter problems with noise and find it essential to smear the heavy-quarks in order to see a signal even at short time separations. Consequently, they are unable to explore whether their data is asymptotic. They have the advantage that the forward recursion is numerically easy and that $v$ is a free parameter, so that many values of $\omega$ can be sampled. The Isgur-Wise function they obtain is consistent with the estimates from lattice QCD and with experiment, within quite large errors, although the renormalisation factor has not yet been included [20]. From a quadratic fit they obtain $\xi'(1) = -0.95$.

It will be important for understanding the size of power corrections to obtain a more precise estimate by this method.

## 4. CONCLUSIONS

The Isgur-Wise function has shown itself to be highly amenable to lattice calculations. Most of the work to date has involved extracting estimates from semi-leptonic decay matrix elements in quenched lattice QCD at around the charm quark mass. Statistical errors appear to be under control and comparable in size to experimental errors. The slope at zero recoil, $-\rho^2$, is not yet precisely determined, but seems to be around $-1.2\pm50\%$. This uncertainty does not greatly affect the extraction of $V_{cb}$ and a value of $0.04\pm20\%$ is obtained by using the lattice measurements to extrapolate experimental data.

For the form factors $h_+$ and $h_{A_1}$, which are protected by Luke's Theorem, there are indications that heavy-quark symmetry holds at heavy-quark masses around charm and light-quark masses around strange. In contrast, large $O(1/m_Q)$ corrections are observed in the form factors $h_-$ and $h_V$, which are not so protected. Lattice HQET offers a good handle on these power corrections by establishing the $m_Q \to \infty$ limit directly.

There is scope for substantial improvements to the analyses reported this year, and the possibility of extending the calculations to semi-leptonic decays of the $\Lambda_b$. Lattice calculations of the Isgur-Wise function are off to an exciting start.